\documentclass[12pt,preprint]{aastex}
\def\gsim{\ \raise 3pt \hbox{$>$} \kern -8.5pt \raise -2pt \hbox{$\sim$}\ }
\def\lsim{\ \raise 3pt \hbox{$<$} \kern -8.5pt \raise -2pt \hbox{$\sim$}\ }
\def\cml{{\rm cm-}$\lambda$}
\def\dml{{\rm dm-}$\lambda$}
\def\ml{{\rm m-}$\lambda$}
\begin{document}
\title{Broadband Quasi-Periodic Radio and X-ray Pulsations in a Solar Flare}
\author{Gregory D.
Fleishman\altaffilmark{1,2}, T. S. Bastian\altaffilmark{3}, and Dale
E. Gary\altaffilmark{1}}

\altaffiltext{1}{New Jersey Institute of Technology, Newark, NJ
07102}  \altaffiltext{2}{Ioffe Physico-Technical Institute, St.
Petersburg 194021, Russia} \altaffiltext{3}{National Radio Astronomy
Observatory, Charlottesville, VA 22903}

\begin{abstract}
We describe microwave and hard X-ray observations of strong
quasiperiodic pulsations from the GOES X1.3 solar flare on 15 June
2003. The radio observations were made jointly by the Owens Valley
Solar Array (OVSA), the Nobeyama Polarimeter  (NoRP), and the
Nobeyama Radioheliograph (NoRH). Hard X-ray observations were made
by the Ramaty High Energy Solar Spectroscopic Imager (RHESSI). Using
Fourier analysis, we study the frequency- and energy-dependent
oscillation periods, differential phase, and modulation amplitudes
of the radio and X-ray pulsations. Focusing on the more complete
radio observations, we also examine the modulation of the degree of
circular polarization and of the radio spectral index.  The observed
properties of the oscillations are compared with those derived from
two simple models for the radio emission. In particular, we
explicitly fit the observed modulation amplitude data to the two
competing models. The first model considers the effects of MHD
oscillations on the radio emission. The second model considers the
quasi-periodic injection of fast electrons. We demonstrate that
quasiperiodic acceleration and injection of fast electrons is the
more likely cause of the quasiperiodic oscillations observed in the
radio and hard X-ray emission, which has important implications for
particle acceleration and transport in the flaring sources.
\end{abstract}

\section{Introduction}

Oscillations and quasi-periodic pulsations (QPPs) have been
observed in the radio emission from solar flares and associated
phenomena for many years (Young et al. 1961). Several types of
pulsation phenomena have been noted. At meter wavelengths (\ml)
certain type IV and moving type IV radio bursts produce
exceptionally regular and deeply modulated pulsations with
bandwidths $\Delta f/f>1$ \citep{Abrami_1970, Abrami_1972,
Rosenberg_1970, Rosenberg_1972, McLean_etal_1971,
McLean_Sheridan_1973, Trottet_etal_1981,Karlicky_Barta_2007}. At
\ml\ and decimeter wavelengths (\dml), rapid QPPs occur with
periods of  10s to 100s of ms. These typically have rapidly
varying amplitudes, variable periods, and bandwidths $\Delta
f/f\lsim 1$ \citep[e.g.,][]{Young_etal_1961, Droge_1967,
Gotwols_1972, Elgaroy_Sveen_1973, Droge_1977, Pick_Trottet_1978,
Bernold_1980, Slottje_1981, Aschwanden_etal_1985,
Aschwanden_Benz_1986, Li_etal_1987, Zlobec_etal_1987,
Stepanov_Yurovsky_1990, Kurths_etal_1991, Yurovsky_1991,
Aschwanden_etal_1995, Fl_etal_2002, PRL, Magdalenic_etal_2002,
Benz_etal_2005, Meszarosova_etal_2005, Chen_Yan_2007}. Finally,
QPPs are observed in microwaves (\cml) with periods $\tau\sim 10
s$, often in association with QPPs in hard X-rays \citep[HXR;
e.g.,][] {Parks_Winckler_1969, Janssens_etal_1973,
Zaitsev_Stepanov_1982b, Kane_etal_1983, Nakajima_etal_1983,
Asai_etal_2001, Grechnev_etal_2003, Nakariakov_etal_2003,
Stepanov_etal_2004, Melnikov_etal_2005}.

A number of models have been proposed to account for the various
types of oscillations and QPPs observed at radio wavelengths
\citep[e.g.,][]{Chiu_1970, Zaitsev_Stepanov_1975a,
Zaitsev_Stepanov_1982a, Nakariakov_etal_2003, Karlicky_Barta_2007},
see also review by \cite{Aschwanden_1987}, and references therein.
Generally speaking, for radio emission attributed to coherent
radiation processes (e.g., fast \dml\ pulsations), QPPs are believed
to result from nonlinear, self-organizing wave-wave or wave-particle
interactions \cite[e.g.,][]{Zaitsev_Stepanov_1975b,
Meerson_etal_1978, Bardakov_Stepanov_1979, Aschwanden_Benz_1988,
Fleishman_etal_1994, Korsakov_Fl_1998}. For radio emission
attributed to incoherent gyrosynchrotron radiation from energetic
electrons (certain type IV and \cml\ emissions) QPPs are believed to
result from modulation of the source parameters (such as  the
energetic electron distribution,  the magnetic field strength, the
line of sight, etc.) via MHD oscillations (kink, sausage, or
torsional modes) and/or modulation of electron acceleration and
injection.

The most recent observations of \cml\ QPPs have been performed with
high spatial resolution in two dimensions by the Nobeyama
Radioheliograph (NoRH) at 17 and 34 GHz (Nakariakov et al. 2003;
Melnikov et al. 2005). In some cases, both \cml\ and hard X-ray
imaging observations have been available (Asai et al. 2001, Grechnev
et al. 2003). These have yielded new insights into the morphology
and the spatial association of the pulsating source(s).
Interestingly, these analyses have all concluded that MHD
oscillations may play a fundamental role in modulating the observed
gyrosynchrotron emission, although the possibility of the
quasiperiodic injection has never been definitely excluded
\citep[see recent review papers,][for further
details]{Nakariakov_Stepanov_2007, Nindos_Aurass_2007}. Here, we
also analyze \cml\ observations of QPPs in a solar flare. In
contrast to the recent studies cited, however, these observations
include excellent   spectroscopic coverage across the \cml\ range as
well as several HXR energy bands observed by RHESSI.  We perform a
Fourier analysis of the radio and X-ray observations of QPPs and
characterize the properties of the pulsations.   In addition, we
consider the spectral index and polarization of the total and
modulated radio emission.  With   this detailed information, we are
able to distinguish between the possible causes of the QPPs. The
data lead us to conclude that the QPPs, at least in this instance,
and possibly more generally, are the result of quasi-periodic
acceleration and injection of electrons rather than MHD
oscillations.

\section{Observations}

The GOES X1.3 solar flare occurred near the limb (S07, E80) in NOAA
active region 10386 on 2003 June 15. In soft X-rays, the flare
commenced at 23:25 UT, and achieved its maximum at 23:56 UT.
Fortuitously, the time was such that the flare could be
simultaneously observed by both the Owens Valley Solar Array (OVSA)
in California, and the Nobeyama Radio Polarimeters (NoRP) and the
Nobeyama Radioheliograph (NoRH) in Japan.  In addition, the flare
was observed in HXRs by the Ramaty High Energy Solar Spectroscopic
Imager (RHESSI). A single SOHO/EIT 195\AA\ image is available during
the flare at 23:46:16 UT.

\subsection{Instrumentation}

The OVSA interferometer \citep{Hurford_etal_1984, Gary_Hurford_1994}
is a solar-dedicated array composed of two 27~m antennas and, at the
time, four 2~m antennas. The array typically observes 40 frequency
channels distributed logarithmically over the frequency range 1 --
18 GHz. The field of view of the 27~m antennas is
$\theta_{27}=46.5/f_9$ arcmin, where $f_9$ is the observing
frequency in GHz. The 27~m antennas therefore resolve the Sun over
most of the observable frequency range and must be pointed to
specific targets of interest on the solar disk. The field of view of
the 2~m antennas is $\theta_2=10.5/f_9$ deg, which does not
significantly resolve the solar disk over the OVSA frequency range.

In the case of the June 15 flare, the array was pointed at AR 10380
(S16W36) and not at AR 10386 so the data from the 27~m antennas are
not useful. We instead rely on observations by the 2~m antennas.
However, due to the fact that the flare occurred late in the
observing day at OVSA, and because the number of baselines is
insufficient to adequately image the source with the 2~m antennas,
we only make use of total power spectra here. Moreover, the antennas
reached their hour angle limit near the end of the flare and the
antennas stopped tracking the target active region (AR 10380)
located at the  western part of the solar disk. Consequently, the
radio source moved in the 2~m field of view so that the array beam
moved towards the radio source as a function of time. Two
corrections to the total power flux as a function of frequency were
made: one to correct for the fact that the antennas were not
pointing at AR 10386, and another to correct for the motion of the
source relative to the 2~m primary beam taper. These corrections
were minor at low frequencies but became important at high
frequencies. The total power data were acquired with a time
resolution of 4~s. No polarization data are available because OVSA
polarimetry at the time required the use of the 27~m antennas.

\begin{figure}
\epsscale{0.85}
\plotone{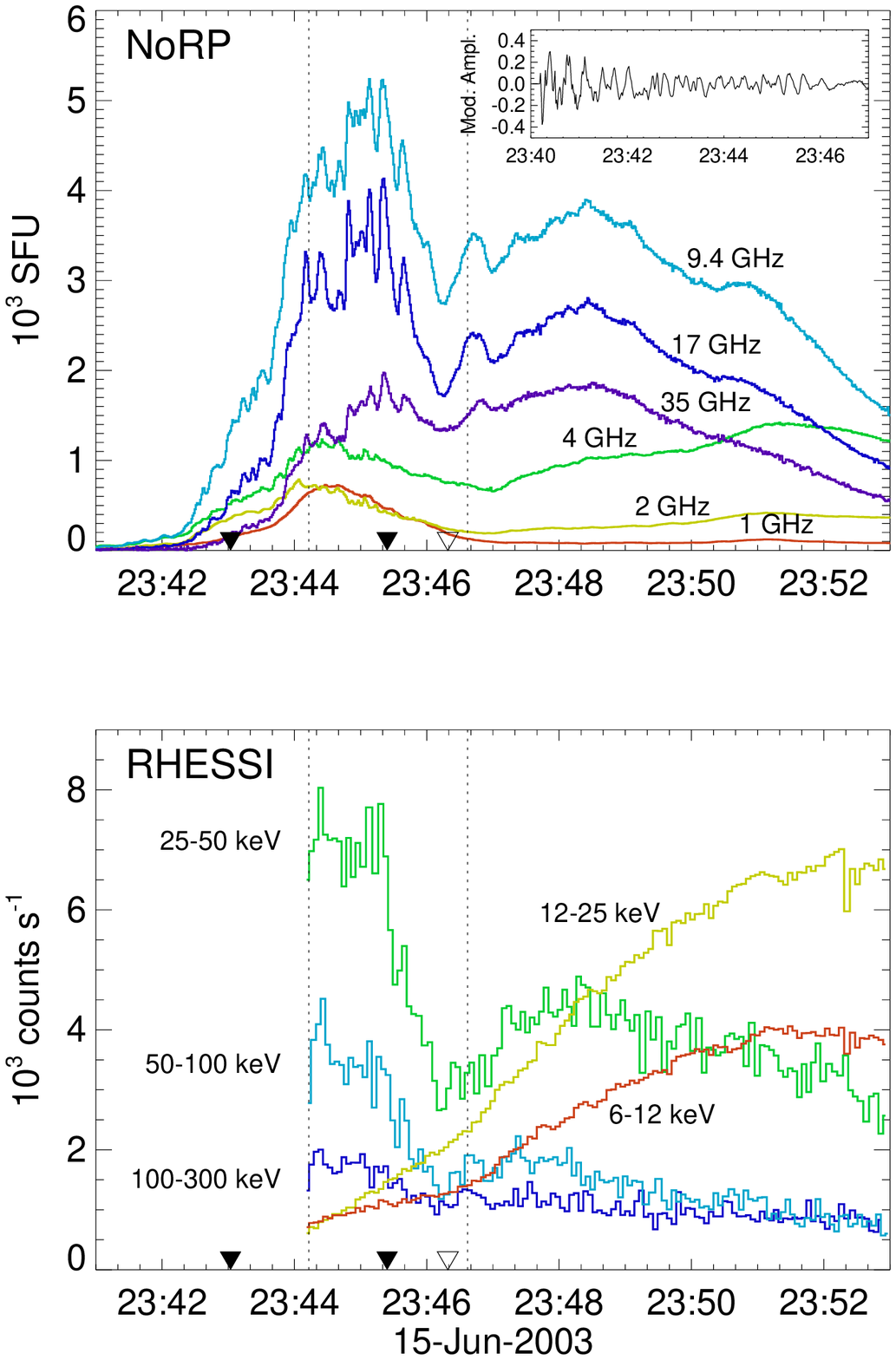} 
\figcaption{\small  Top: NoRP light curves. The normalized
modulation at 9.4 GHz is shown in the inset. The inverted triangles
indicate the times shown in Figure~\ref{summary}, the open triangle
indicating the time of the EIT 195\AA\ image. Bottom: RHESSI light
curves for the same time range. The vertical dashed lines indicate
the time range selected for analysis. \label{overview}}
\end{figure}

The NoRP \citep{Torii_etal_1979, Nakajima_etal_1985} provides total
power data in total and circularly polarized intensity (Stokes
parameters I and V) at 1, 2, 3.75, 9.4, 17, 35, and 80 GHz with a
time resolution as high as 0.1~s. For the purposes of the analysis
presented here, the data were averaged to a time resolution of 1~s.
The NoRP data broaden the spectral range covered by the radio
observations and provide a useful check against discrete OVSA
frequency channels as well as constraints on the polarization of the
observed emission.  In the present case the quality of the 80~GHz
data were insufficient for quantitative analysis.  An overview of
the NoRP observations is shown in the top panel of
Figure~\ref{overview}.

The flare was also observed by the NoRH \citep{Nakajima_etal_1994},
which images the Sun at 17 and 34 GHz with a time resolution as high
as 0.1 s. Again, in the present case, the data were averaged to a
time resolution of 1~s. A time series of maps was then created in
each frequency band spanning the duration of the flare. The maps
were created using the AIPS software package. The AIPS task IMAGR
was used to produce each map and to deconvolve the point spread
function of the NoRH (the "dirty beam"). The snapshot data were
uniformly weighted, which enhances the weight given to the long
antennas baselines relative to the short baselines, thereby
improving the angular resolution with which the radio emission can
be imaged. In the present case, the angular resolution of the maps
is $15.4"\times 12.8"$ and $7.8"\times 7.1"$ at 17 and 34 GHz,
respectively.

Data from RHESSI \citep{Lin_etal_2002} are available beginning at
approximately 22:44 UT. In fact, RHESSI was off-pointed from the Sun
to observe the Crab nebula during June 14-15 and the flare was 1.2
to 1.5 degrees off axis, well outside the normal field of view (G.
Hurford 2006, private communication). HXR photon counts were
nevertheless accumulated during each half rotation of the spacecraft
when its collimating grids were favorably oriented toward the flare.
The data were corrected for the varying off-axis grid transmission
and detector illumination and converted to one corrected count rate
per spacecraft rotation.   HXR light curves were obtained in five
energy bands (6-12 keV, 12-25 keV, 25-50 keV, 50-100 keV, and
100-300 keV) starting at 23:42:11  UT with a time resolution of
approximately 4~s. We have not imaged the data, nor do we rely on
the HXR data for spectroscopic analysis. We only make use of
quantities that are independent of calibration, as described further
in the next section. An overview of the RHESSI data is shown in the
bottom panel of Figure~\ref{overview}.

To summarize, our data set includes 1-18 GHz total flux density
(Stokes I) spectra from OVSA, measurements of total flux and
circularly polarized flux (Stokes V) from NoRP, imaging measurements
at 17 (I/V) and 34 (I) GHz from NoRH, and HXR light curves in five
photon energy ranges from RHESSI. We now examine the properties of
the pulsations in the radio and HXR wavelength regimes.

\subsection{Data Analysis and Results}

\subsubsection{Radio Mapping}

\begin{figure}
\epsscale{0.7}
\plotone{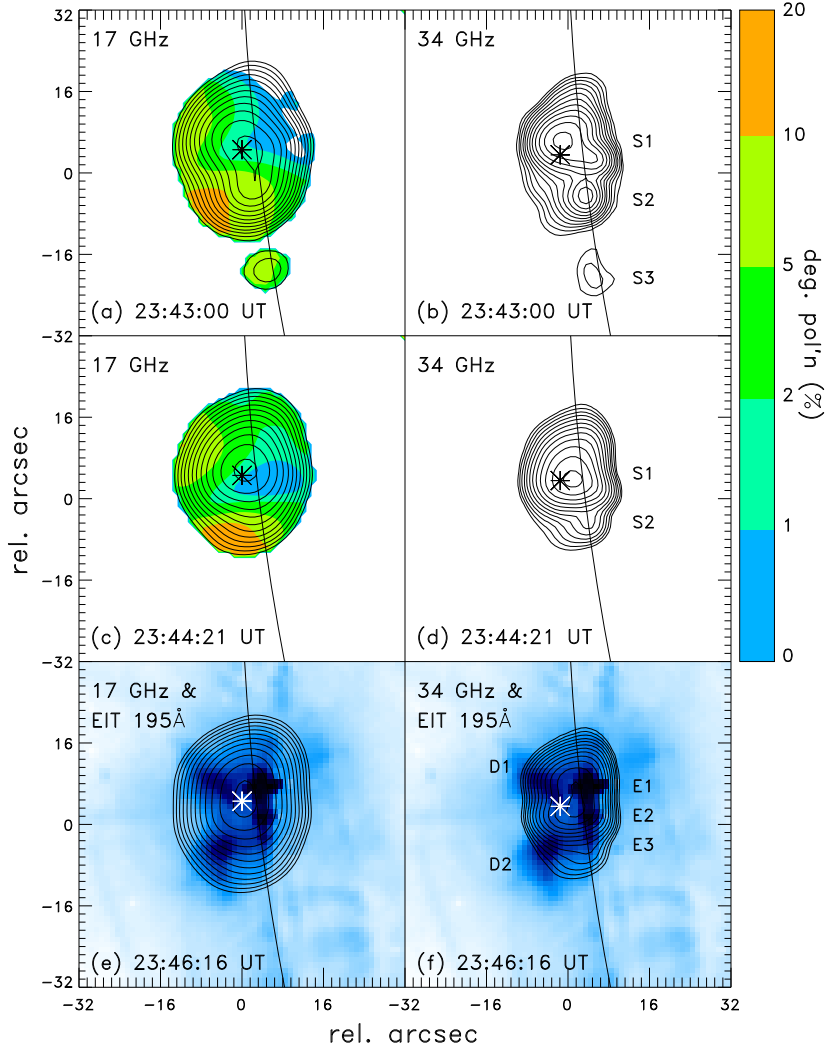} 
\figcaption{\small  Overview of the radio source imaged by the NoRH
at the times indicated by inverted triangles in
Figure~\ref{overview}. The solid arc indicates the solar limb. Radio
contours are at intervals of $2^{(n+1)/2}$ percent of the peak
brightness temperature of each map, where $n=1, 2, ... 12$. Color
contours show the degree of polarization at 17 GHz. Panels (a) and
(b) show contour maps of the 17 and 34 GHz source during the rise
phase, when the peak brightness temperatures are $3.4\times 10^7$ K
and $3.2\times 10^6$ K, respectively; panels (c) and (d) show the
same at the time of flux maximum, when the peak brightness
temperatures are $3.0\times 10^8$ K and $5.5\times 10^7$ K,
respectively; panels (e) and (f) show the same superposed on the EIT
195\AA. The brightness temperature maxima are $1.0\times 10^8$ K and
$3.0\times 10^7$ K, respectively. The asterisk in each case shows
the mean position of the quasi-periodic pulsations.
 \label{summary}}
\end{figure}

The NoRH imaging observations at 17 and 34 GHz are summarized in
Figure~\ref{summary} where contour maps in each frequency are shown
during the rise phase (Figure~\ref{summary}a,b), the flux maximum
(Figure~\ref{summary}c,d), and near the end of the first major peak
of the flare when a single image from SOHO EIT at 195\AA\ is
available (Figure~\ref{summary}e,f). The corresponding times are
indicated by inverted triangles in Figure~\ref{overview}. The peak
brightness temperatures at each time and in each band are given in
the figure caption. While a time series of such maps is available in
each frequency band with a time resolution of 1~s, it is neither
necessary nor practical to reproduce them all here.

The FWHM angular size of the 17~GHz source varies between
$\approx\!22"\times 25"$ and $\approx\!25"\times 33"$ during the
course of the flare, while the 34 GHz source varies between
$\approx\!17"\times 19"$ and $\approx\!19"\times 22"$. The source is
not well-resolved at 17~GHz since the angular resolution is
$15.4"\times 12.8"$. The angular resolution at 34 GHz is $7.8"\times
7.1"$, however, and the source is better resolved.
Figure~\ref{summary}b shows the 34~GHz source at 23:43:00~UT; it is
composed of three components labeled S1, S2, and S3. Source S3
remains very faint during the course of the event and is not visible
in subsequent images. Source S1 may correspond to a coronal loop
structure, whereas source S2 is compact and unresolved. S1 is the
dominant source in all contour maps shown in Figure~\ref{summary}
and, indeed, is the dominant source in all maps over the time range
indicated by the vertical dashed lines in Figure~\ref{overview}. As
we show in \S2.2.3, it is also the source of the QPPs.

The EIT~195\AA\ image obtained at 23:46:16~UT is complex, showing three compact bright patches inside the solar limb (labeled E1, E2, and E3); E1 and E2 are slightly saturated. Above the limb, two diffuse patches of emission are seen. It is possible that sources E1, E2, D1, and S1 are associated with a single loop or loop system, with E1 and E2 marking the footpoints. Sources  E3, D2, and S2 may be associated with a separate loop or loop system that extends to the southeast. We note that soft X-ray images from the GOES 12 Soft-X-ray Imager (SXI) are also available. We find that at the time of radio maximum the SXR emission appears to originate from a single diffuse source, similar in appearance and location to the 17 GHz source. For this reason, we do not reproduce the SXI data here. The SXI data are consistent with our identification of S1 being the dominant source during the first major peak of the flare.

Maps of Stokes V are available from the NoRH at 17 GHz, and shown by
the color contour map in Figure~\ref{summary}a and c. We find that
the angular resolution at 17 GHz is insufficient to enable us to use
the Stokes V maps as a basis to draw any firm conclusions regarding
source morphology. We also note that since the source is near the
limb, the aspect angle is far from optimum for this purpose. While
we are unable to draw definite conclusions regarding the detailed
source morphology in the 17 and 34 GHz bands, we can conclude that
the flux density of the radio source is dominated by S1 in each
band. Based on its morphology at 34 GHz (Figure~\ref{summary}b) we
suggest that S1 may be a coronal loop.

We now turn to a more quantitative analysis of the radio and HXR
QPPs. We return to the question of the source morphology in \S2.2.3.

\subsubsection{Modulation Amplitudes}

In order to characterize the QPPs in the radio bands, we calculate
the {\sl modulation power} $P(f)$ and the {\sl modulation amplitude}
$m(f)=\sqrt{P(f)}$ for each frequency:

\begin{equation}
\label{mod_ampl_def}
 P(f)  = \left< S^2(f,t)\right> = \frac{1}{T} \int\limits_0^T S^2(f,t) dt,
\end{equation}
where
\begin{equation}
\label{Osc_comp_def}
 S(f,t) = \frac{F(f,t)-\left< F(f,t) \right>}{\left< F(f,t) \right>}
\end{equation}
is the normalized modulation of the signal, $f$ is the frequency in
GHz, $t$ is the time in seconds, $F(f,t)$ is the total flux density
at frequency $f$. Similarly, a modulation amplitude can be defined
for the RHESSI HXR bands, with $F(f,t)$ replaced by $C(E,t)$, where
$C$ represents the photon counts s$^{-1}$ in a given energy band
$E$. The brackets denote a running average of the original signal
over time. The modulation amplitude is a measure of the variation of
the emission with respect to the running average.  While the
averaging time used here for the running average is 20~s, the
analysis is insensitive to the precise value used. The inset to
Figure~\ref{overview} shows the normalized modulation at 9.4 GHz
from the beginning of the impulsive phase at 23:40 UT through the
first major peak. We note the clear decline in modulation amplitude
from the onset of the emission, when the modulation is $\gsim 30\%$
of the mean flux, to the time of the flare maximum and later when
the
modulation amplitude is $\approx 10\%$ of the mean. 
Figure~\ref{modDepth} displays the spectral dependence of the radio
and HXR modulation amplitudes calculated for the time period
23:44:11-23:46:35 UT.

\subsubsection{Fourier Analysis}

The use of Fourier analysis to study radio pulsations is described
by \cite{Fl_etal_2002, PRL}, who used it to study millisecond
pulsations of coherent radio emission, and by
\cite{Melnikov_etal_2005}, who used it to study QPPs from a flare
observed by the NoRH. We apply and extend that approach here,
considering both multi-frequency radio and HXR data. We confine our
attention to the time range 23:44:11-23:46:35 UT demarcated in
Figure~\ref{overview} by vertical dotted lines.  The start time is
determined by the availability of RHESSI data. The end time
corresponds to the end of the first major peak of the \cml\ and HXR
emission (Figure~\ref{overview}). The results obtained for the radio
emission were nevertheless checked and found to be stable against
shifts in the starting and/or ending time of the interval analyzed.

The power spectrum and phase were computed from the Fourier
transform of the normalized modulation for each frequency in the
OVSA and NoRP data. The normalized modulation and the corresponding
power spectrum are displayed for each frequency observed by the NoRP
and for the nearest corresponding OVSA frequency in
Figure~\ref{Fig3}. We note the presence of several peaks in each
spectrum. The two most significant peaks at the optically thin
frequencies of 17 and 35 GHz correspond to power at periods of
$\approx\!14.5$~s and $\approx\!18.4$~ s. The most significant peak
at lower (optically thick) frequencies corresponds to a period near
21~s.

Figure~\ref{Fourier_radio} shows a more comprehensive display of the
radio oscillations in total power and their Fourier amplitudes and
phases. Panel (a) shows the dynamic spectrum of the total flux
density between 1-18 GHz observed by OVSA, while panel (b) shows the
normalized modulation as a function of time and frequency. Panel (c)
shows the Fourier amplitude of the normalized modulation as a
function of the observing frequency and oscillation frequency for
the time range selected for analysis. The two main amplitude peaks
and subsidiary peaks are clearly seen. The corresponding phase of
these peaks is shown in panel (d), indicating phase coherence over
the frequency range considered. In agreement with the NoRP data, the
OVSA data show a systematic change in the oscillation frequency of
the dominant peak from approximately 18.4 to 21 s between 18 and 1
GHz. The phase and the corresponding time delays are shown as a
function of frequency in Figure~\ref{phase_delay} for both the NoRP
and OVSA data. We find that the relative phase and delay is such
that lower frequencies lead higher frequencies. We note that a
standard cross-correlation analysis reveals the same trends as
Fourier analysis does.

We now   return to  the radio imaging observations obtained by the
NoRH at 17 and 34 GHz to localize the primary source of the radio
pulsations. We have Fourier analyzed the time series of maps in both
frequency bands on a pixel by pixel basis. In this case, we are
interested in which pixels dominate the net QPP amplitude (or power)
and we therefore Fourier analyze $F(x,y,f,t)-\langle
F(x,y,f,t)\rangle$, where the brackets again denote the running time
average, for all pixels $(x,y)$ at $f=17$ and 34 GHz for the time
range   23:44:11--23:46:35~UT. The result is maps of the Fourier
amplitude (or power) and phase for each frequency $\tau_P^{-1}$
contributing to the QPPs at each spatial location. We find that the
17 and 34 GHz power spectra at the location of S1 in
Figure~\ref{summary} look remarkably similar to those shown for the
NoRP total power measurements made at 17 and 35 GHz, shown in
Figure~\ref{Fig3}.
\par
Consider the two dominant peaks in the power spectra shown in
Figure~\ref{Fig3} at 9.4, 17, and 35 GHz, at frequencies
corresponding to periods of 14.5 and 18.8~s. We show maps of the
Fourier amplitude and phase at these two frequencies in
Figure~\ref{norh_phases}.  The black contours represent the Fourier
amplitude and the asterisk indicates the location of the amplitude
maximum. The color represents the phase relative to that at the
location of the asterisk in each case  with dark blue representing
$-180^o$ and dark red representing $+180^o$ phase shifts. The white
contours show the flux density at the time of flux maximum and are
identical to those show in Figure~\ref{summary}c,d.
Figure~\ref{norh_phases} shows that the peaks at periods of 14.5 and
18.4~s are essentially co-located at both 17 and 34 GHz, that their
amplitude maxima coincide with the location of S1 in
Figure~\ref{summary}, and that they are spatially phase coherent;
that is, the phase of each peak does not vary appreciably in the
vicinity of the amplitude maximum in each case. We conclude that not
only is the total flux at 17 and 34/35 GHz   dominated by S1, the
QPPs originate in S1.

Turning now to the HXR data, we also Fourier analyzed each of the
five RHESSI energy ranges. Figure~\ref{Fig4} shows the HXR
normalized modulations and the corresponding power spectra for each
energy range. The 17 GHz normalized modulation and power spectrum is
over-plotted on  four energy bands showing strongest modulations in
order to compare the radio and HXR spectra. It is clear that the two
most prominent peaks in the radio spectra are present in the HXR
spectra, too, as well as some subsidiary peaks. The frequencies of
the two main HXR peaks correspond to periods of 15~s and 19~s,
similar but marginally somewhat longer than the two dominant periods
in the optically thin radio emission.
We again compute the (relative) phase and delay for each energy
range (Figure~\ref{X_fft}). We find that the higher energy ranges
tend to be progressively delayed relative to the lower energy
channels.

In the remaining observational subsections we direct our attention
to total power measurements of the radio emission, as observed by
OVSA and NoRP, and the HXR count rates as observed by RHESSI. The
optically thin radio flux density is clearly dominated by S1, as are
the QPPs. Given that the dominant peak in the QPP spectra near 20~s
persists at all radio frequencies between roughly 1-35 GHz in a
phase coherent fashion, we conclude that the QPP source is likely
localized to S1 at all frequencies. While the location of the HXR
source or sources is unknown, given the similarity between the radio
and HXR QPPs, they are   likely intimately related, both temporally,
and spatially.

\subsubsection{Partial modulation amplitudes}

Besides the above, the Fourier analysis allows consideration of the
total modulation power (\ref{mod_ampl_def}) and the partial
modulation of the signal by each pulsating component described by a
Fourier peak with a finite bandwidth. The need to consider the
partial modulation amplitude arises because more than one
significant Fourier peak is present in the power spectrum.  It is
not immediately clear whether the same, or different, physical
processes are responsible for each peak, e.g., some might be the
result of quasiperiodic injection of fast electrons, others of MHD
oscillations.

\begin{figure}
\epsscale{0.85} \plottwo{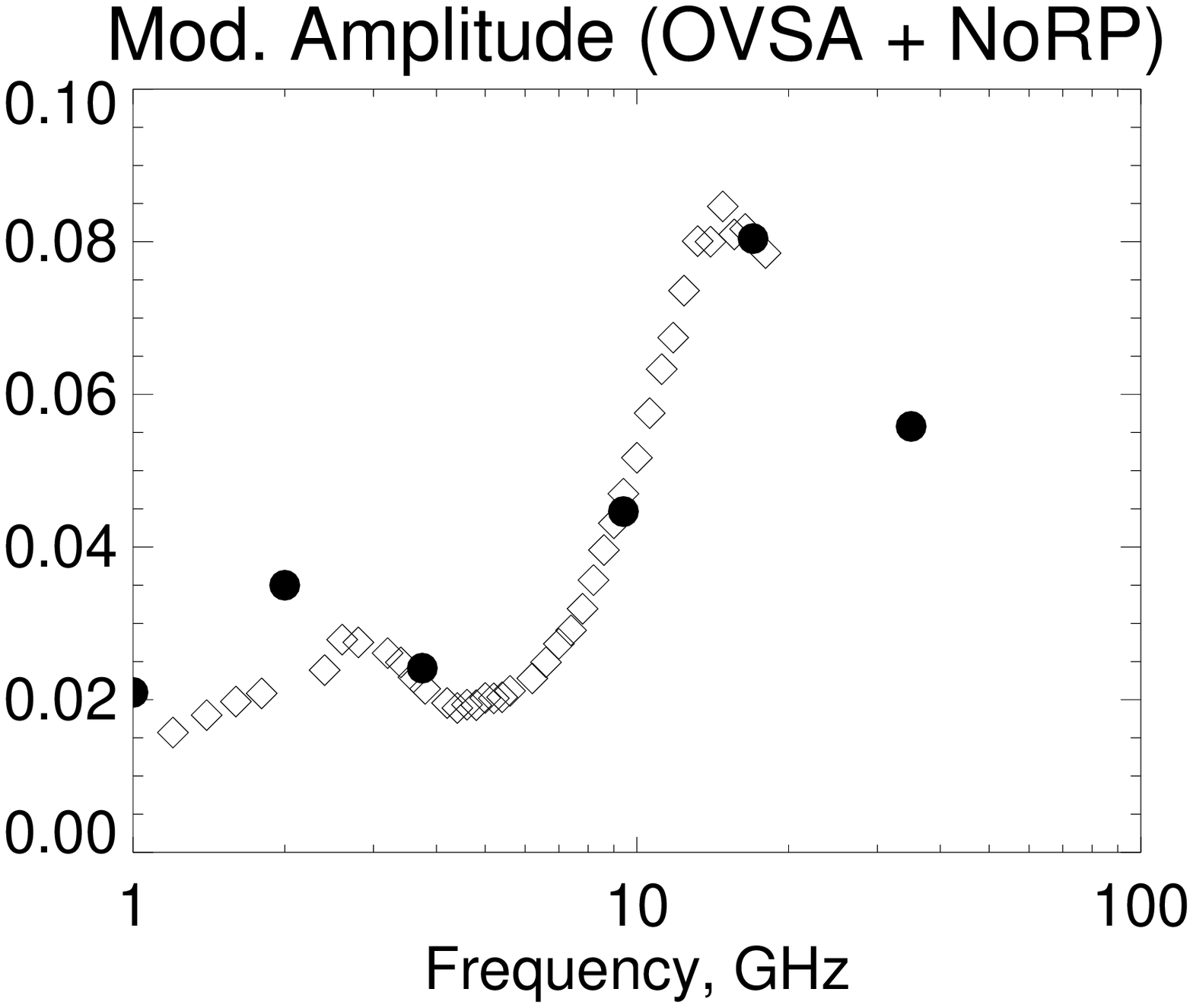}{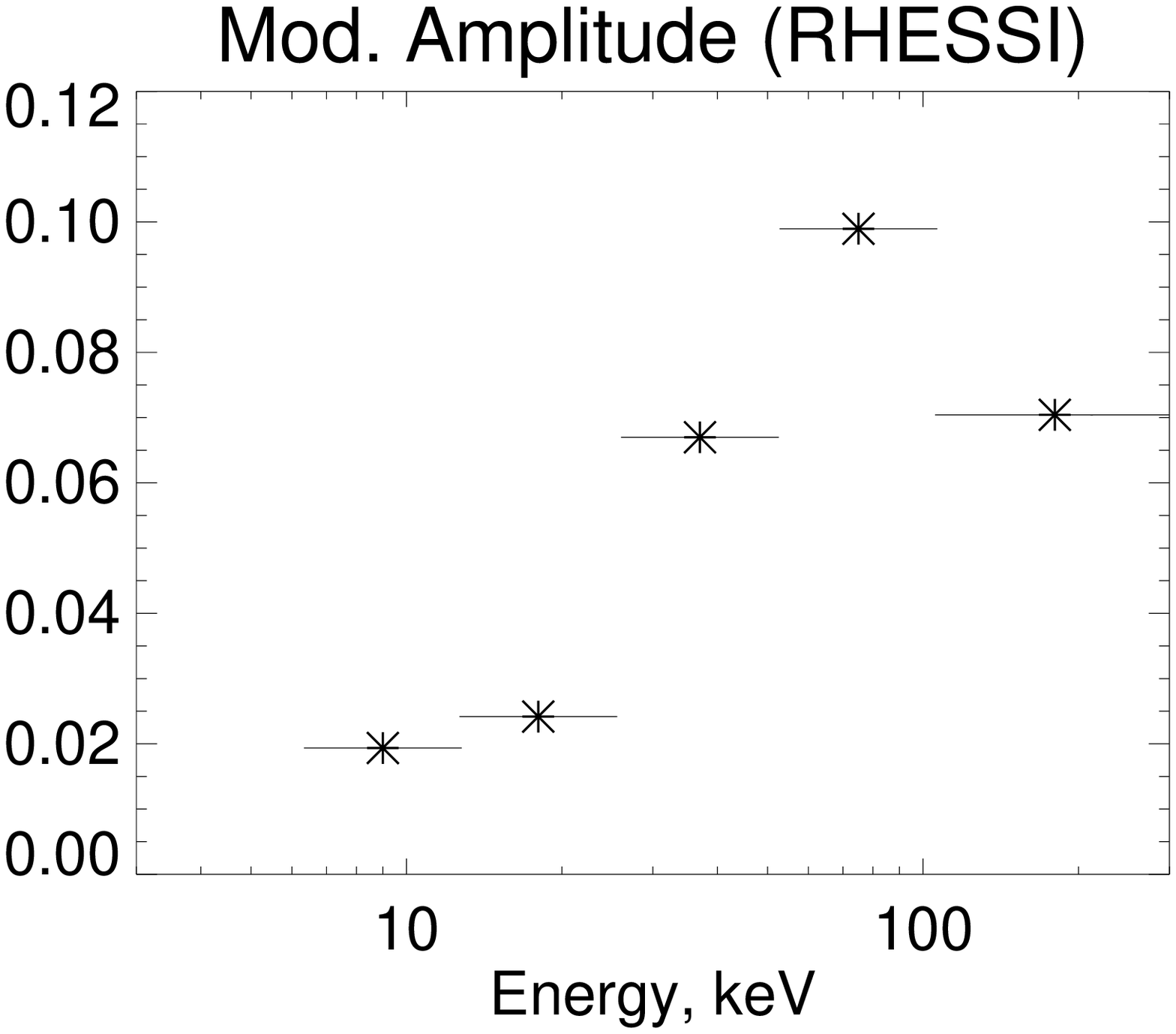}
 \figcaption{\small  The
modulation amplitude of the radio emission (left) obtained with OVSA
(diamonds in both this and further figures) and NoRP (filled circles
in both this and further figures) calculated with
Eq.(\ref{mod_ampl_def}) for the selected time interval and of X-ray
emission (asterisks, right in both this and further figures).
\label{modDepth}}
\end{figure}

\begin{figure}
\includegraphics[angle=0,scale=0.85]{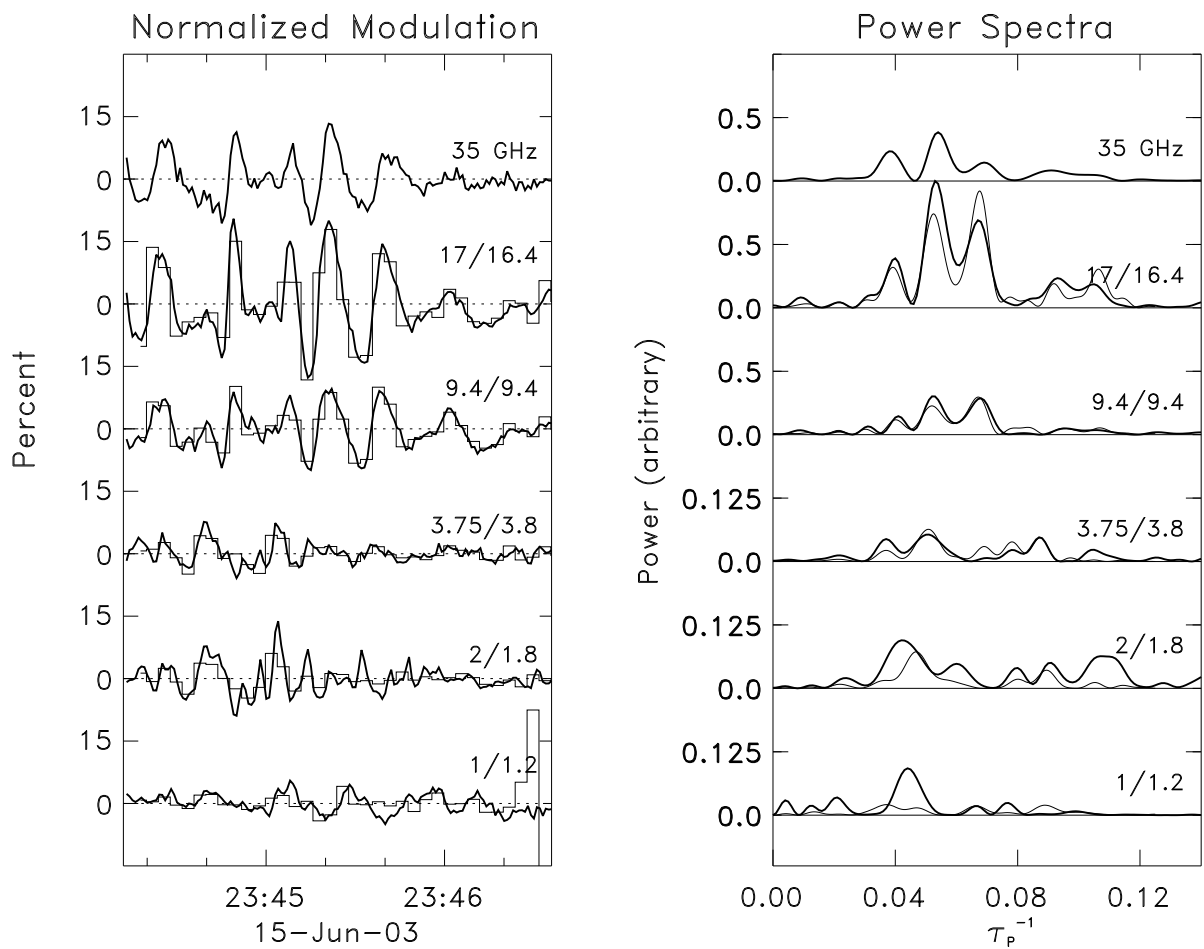}
\figcaption{\small  The normalized modulation of the NoRP and OVSA
data (left) and the corresponding power spectra (right). The two
most prominent peaks in the top three panels of the power spectra
correspond to periods $\tau_P$ of approximately 19~s and 15~s. A
lower-amplitude peak appears at a period of approximately 10~s. The
lower three panels (note the change of scale) display a common peak
near a period $\tau_P\approx 21$~s. \label{Fig3}}
\end{figure}

\begin{figure}
 \epsscale{0.75} \plotone{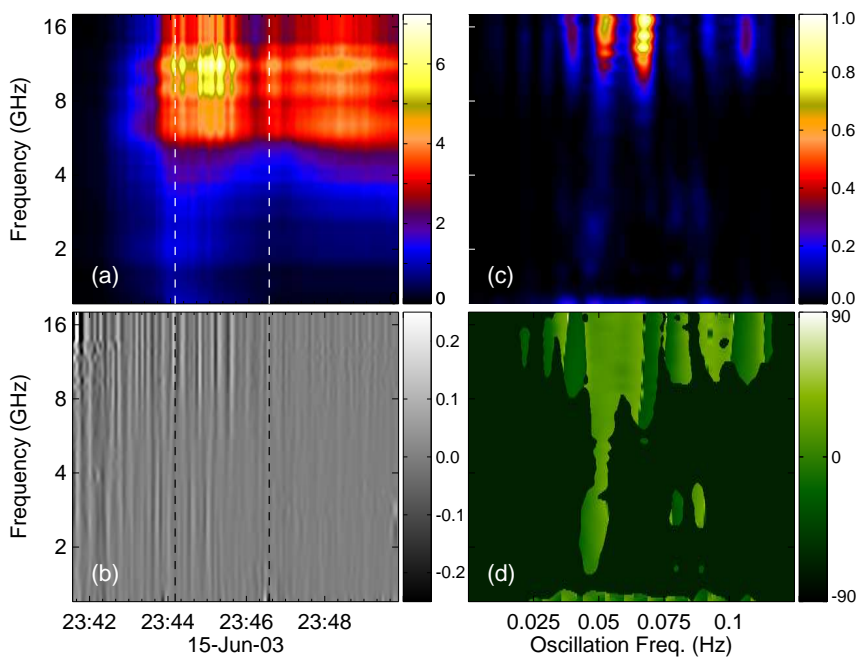}
 \figcaption{\small  a) OVSA dynamic spectrum of the 1-18 GHz total intensity. The
units of the intensity wedge to the right of the panel are 1000 SFU;
b) the normalized modulation of the 1-18 GHz radio emission as a
function of time and frequency. The scale ranges from $-25\%$ to
+25\%; c) the Fourier amplitude of the normalized modulation as a
function of oscillation frequency and radio frequency for the time
interval shown between the vertical dashed lines in panels a and b.
The maximum amplitude has been scaled to 1.0; d) the Fourier phase
of the normalized modulation as a function of oscillation frequency
and radio frequency, again for the time range shown by the vertical
dashed lines in panels a and b. The phases are scaled from
$-90^\circ$ to $+90^\circ$. \label{Fourier_radio}}
\end{figure}

To introduce the partial modulation amplitudes we apply a
fundamental property of the Fourier transform, i.e., Parceval's
identity:
\begin{equation}
\label{FFT_equiv}
 \sum_{i=0}^{N_i-1} S^2(f,t(i)) \equiv
N_{tot}  \sum_{n=0}^{N_{tot}-1} \mid S(f,\nu(n)) \mid^2
  ,
\end{equation}
where $N_i$ is the total number of the measurements in the time
domain and $N_{tot}$ is the total number of the oscillation
frequencies $\nu(n)$ where the Fourier amplitudes are determined.
Accordingly, combining this equivalence with the definition of the
total modulation power (\ref{mod_ampl_def}), we can express the
modulation amplitude as a sum over all Fourier harmonics:
\begin{equation}
\label{mod_depth_FFT}
  m(f)  =
\left(\frac{N_{tot}}{N_i} \sum_{n=0}^{N_{tot}-1} \mid S(f,\nu(n))
\mid^2\right)^{1/2}
  .
\end{equation}
The total modulation amplitude, $m(f)$, related to contribution of
all available Fourier harmonics is given in Figure~\ref{modDepth}.
Here we consider the partial modulation amplitudes, $m_p(f)$,
related to modulation of the radio emission by a limited region of
Fourier harmonics from $n_1$ to $n_2$ covering each of the main
Fourier peaks, defined as follows:
\begin{equation}
\label{mod_depth_partial}
 m_p(f)  =
\left(2\frac{N_{tot}}{N_i} \sum_{n=n_1}^{n_2} \mid S(f,\nu(n))
\mid^2\right)^{1/2}
  ,
\end{equation}
where the factor 2 is related to the fact that each partial
oscillation consists of two Fourier peaks with identical intensities
and opposite phases.

Figure~\ref{periods_radio} displays the frequency dependence of the
partial modulation amplitudes, corresponding to the two main Fourier
peaks. Remarkably, the partial modulation amplitudes for all
significant Fourier peaks determined with OVSA and NoRP data display
similar spectral behavior, i.e., they look qualitatively similar to
each other and to the total modulation amplitude presented in
Figure~\ref{modDepth}. We conclude that all of the Fourier peaks are
due to the same physical process.

\subsubsection{Spectral Index Variations}

Direct measurement of the brightness temperature of $3\times 10^8$ K
at 17 GHz by NoRH at the time of flux maximum suggests that the
corresponding brightness temperature at the spectral peak (around 10
GHz) exceeded $2\times 10^9$ K. This high value indicates that the
spectral peak in our radio burst is formed due to the
gyrosynchrotron self-absorption, rather than by Razin suppression.
The observed radio emission spectrum from a flare is often
characterized by a power law $F(f)\propto f^{-\beta}$, where $\beta$
is the spectral index \citep[e.g.,][]{BBG}. For those frequencies
where the emission is optically thin, $\beta>0$, we adopt
$\beta_{\rm thin}=\beta$, and for those frequencies where the
emission is optically thick, $\beta<0$, we adopt $\beta_{\rm
thick}=-\beta$, so that both indices are positive. The radio spectra
from OVSA and NoRP indicate that the spectral maximum of the source
is $\lesssim\!10$ GHz. We have therefore formed $\beta_{\rm thin}$
using the NoRP 17 and 35 GHz observations, which are optically thin,
and $\beta_{\rm thick}$ using the NoRP 3.75 and 9.4 GHz
observations, which are optically thick (in fact, the 9.4 GHz
emission is probably only partly optically thick, which results in a
spectral index that is somewhat smaller than the true optically
thick spectral index). Figure~\ref{NoRP_curves}a shows the two
spectral indices and their modulation in time. The solid line shows
$\beta_{\rm thin}$ whereas the dashed line shows $\beta_{\rm thick}$
for the time range analyzed throughout this paper. $\beta_{\rm
thin}$ declines systematically with time (i.e., the optically thin
radio spectrum hardens) while $\beta_{\rm thick}$ remains roughly
constant, but both show fluctuations similar to those seen in the
radio time profile. Figure~\ref{NoRP_curves}b shows the normalized
modulation of $\beta_{\rm thin}$ and compares it with that of the 17
GHz emission.  We note that the two are highly correlated. When the
17 GHz total flux is high, the spectral index is larger and the
spectrum is therefore steeper (softer) and vise versa.  Such a
spectral behavior seems to be opposite to the standard
soft-hard-soft evolution observed in individual HXR peaks
\citep{Benz_etal_2005}. Figure~\ref{NoRP_curves}c shows the same for
$\beta_{\rm thick}$ and the 9.4 GHz total flux. The correlation is
again excellent.

\subsubsection{Polarization Variations}

Finally, we consider the polarized radio emission. The degree of
circular polarization, defined as $\rho_c(f)=V(f)/I(f)$, is plotted
for 9.4 and 17 GHz in Figure~\ref{Pol_osc}a. The degree of
polarization is quite low: only $\approx 1.5\%$ at 9.4 GHz and
$5-10\%$ at 17 GHz, the latter showing a systematic decrease with
time. Nevertheless, given the excellent signal to noise ratio, the
normalized modulation  of $\rho_c$ can be formed for both
frequencies. Figure~\ref{Pol_osc}b compares the normalized
oscillating component of $\rho_\nu$ at 9.4 GHz with that of the 9.4
GHz total flux. Figure~\ref{Pol_osc}c presents the same for 17 GHz.
We find that the degree of polarization is anti-correlated with the
radio QPPs in both cases. In other words, when the radio intensity
is high, the degree of polarization is lower and when the radio
intensity is low, the degree of polarization is greater.

\subsection{Summary of Results}

It is useful to summarize the key results from our analysis:

\begin{enumerate}
 \item The radio and HXR flare emission display QPPs that are
highly correlated with each other.
 \item The radio and HXR QPPs
are characterized by several significant peaks in their power
spectra. The two dominant peaks correspond to periods near 15 and
20 s.
\item  Both the radio flux density and the QPPs are dominated by source S1,
shown in Figures~\ref{summary} and \ref{norh_phases}.
 \item Each of the Fourier peaks is due to the same physical
process, as indicated by our partial modulation analysis.
 \item The relative phase of each dominant peak shows a small,
systematic shift as a function of radio frequency. The sense of
the phase shift is such that the low radio frequencies lead the
higher radio frequencies. A similar behavior is observed in the
phase of the two dominant HXR peaks although here, it is the lower
energy bands that lead the higher energy bands.
 \item The total modulation power displays a systematic trend with
frequency   in the radio domain, the maximum modulation occurring at
around 15 GHz. That is, the maximum power of the QPPs occurs
somewhat above the spectral turnover.
 \item Variation in the spectral index of the optically thin radio
emission is correlated with variations in the radio emission. The
sense of the correlation is such that the radio spectrum is steeper
during QPP peaks. This is opposite from the usual soft-hard-soft
behavior for X-ray peaks.
 \item Variation in the degree of circular polarization, $\rho_c$,
is anticorrelated with that in the radio flux at 9.4 and 17 GHz;
i.e., $\rho_c$ is lowest during QPP peaks.
\end{enumerate}

\noindent In light of the above findings, we now consider whether
the observed QPPs are the result of MHD loop oscillations in the
source, or whether they are more consistent with the quasi-periodic
injection of energetic electrons into the source. Unlike many other
studies, the whole body of the data available for the event under
analysis allows making a firm conclusion about this.

\section{Data interpretation}

Electrons are accelerated and injected into coronal magnetic loops
during flares. Due to the fact that coronal magnetic loops have a
weaker magnetic field at the loop top than at their foot points,
they act as magnetic traps.  Whether an electron is trapped or not
is determined by its pitch angle. The critical angle that determines
precipitating vs. trapped electrons is given by $\alpha_{\rm
lc}=\arcsin(B_{\rm inj}/B_{\rm loss})$, where $B_{\rm inj}$ is the
magnetic field strength where the electrons are injected and $B_{\rm
loss}$ is the magnetic field strength at the chromospheric height in
the foot point where the electrons are collisionally removed from
the trap. Electrons with sufficiently large pitch angles
$\alpha>\alpha_{\rm lc}$ are trapped in the magnetic loop until they
are scattered, by Coulomb collisions or wave-particle interactions,
to sufficiently small pitch angles $\alpha<\alpha_{\rm lc}$, or are
thermalized due to \textit{in situ} Coulomb collisions. Electrons
with pitch angles $\alpha<\alpha_{\rm lc}$ immediately propagate to
the chromosphere, where they collide with cold ions and produce HXRs
via nonthermal bremsstrahlung. Both trapped and precipitating
electrons emit radio waves via nonthermal gyrosynchrotron emission.
This basic picture is referred to as the "trap plus precipitation"
\citep[TPP;][]{Melrose_Brown_1976} or the "direct precipitation and
trap plus precipitation" \citep[DPTPP;][]{Aschwanden_1998,
Aschwanden_etal_1998, Aschwanden_etal_1999} model.

\subsection{QPPs as the Result of MHD Oscillations}

As discussed in \S1, radio and HXR QPPs have been interpreted in
terms of variations in the magnetic field and/or variations in the
electron distribution function due to acceleration and injection.
Magnetic field variations are generally attributed to MHD
oscillations. A number of recent studies of \cml\ and HXR QPPs have
concluded that MHD oscillations may be the cause  (Asai et al. 2001;
Grechnev et al. 2003; Nakariakov et al. 2003; Melnikov et al. 2005).
Among the possible oscillation modes considered were the (standing)
sausage, kink, and torsional modes. Grechnev et al., studying the
same event as Asai et al. (1998 November 10), also consider the
possibility of magnetic field variations caused by the launch of
impulsively generated, propagating MHD waves
\citep{Roberts_etal_1984} and conclude that the QPPs studied may
result from either propagating MHD waves or torsional oscillations,
although quasi-periodic modulation of the electron acceleration
could not be definitively excluded due to lack of spectral
information. Nakariakov et al. and Melnikov et al. both suggest that
the QPP event of 2000 January 12 can be accounted for in terms of
global sausage mode oscillations.

Can the QPPs observed from the 2003 June 15 flare be understood in
terms of MHD oscillations? Certainly, the dominant periods in the
oscillations are not inconsistent with possible MHD modes. As noted
in \S2.2, several peaks are present in the power spectra. The two
strongest peaks in the radio and HXR emission have periods near 15
and 20 s. Previous authors have shown that these periods are
consistent with sausage mode oscillations, fast-mode waves, and
torsional oscillations (e.g., Grechnev et al. 2003). However, we can
exclude each of these possibilities in the present case through
consideration of the radio spectroscopic data and its polarization
properties.

First, a sausage mode oscillation or the launch of fast-mode MHD
waves results in a periodic compression and expansion of the
magnetic flux tube in which the radiating electrons reside. Magnetic
flux conservation requires a corresponding variation in the magnetic
field strength and hence, a modulation of the gyrosynchrotron
emissivity and the absorption coefficient. Although, the source
volume and the number density of the fast electrons will be
oscillating, their product will stay constant. An immediate
consequence of a sausage mode oscillation is a $180^\circ$ phase
difference between the optically thick and optically thin emission.
To see this, we take the nonthermal distribution of electrons to be
a power law, with $n_{nt}(E)dE \propto N E^{-\delta}dE$.
\cite{Dulk_1985} gives approximate expressions for the
gyrosynchrotron emissivity $\eta_f$, absorption coefficient
$\kappa_f$, effective temperature $T_{eff}$, and the degree of
circular polarization $\rho_c$, for a power law distribution of
electrons that usefully illustrate the parametric dependences. For
optically thin emission at a fixed frequency the radio emission
$F(f) \sim \eta_f \propto N\ B\
s^{1.22-0.9\delta}(\sin\theta)^{-0.43+0.65\delta}$, where
$s=f/f_{Be}$ is the harmonics of the electron gyrofrequency
$f_{Be}=eB/2\pi m_e c$; $N$ is the total number of nonthermal
electrons above some threshold energy $E_\circ$; and $\theta$ is the
angle between the magnetic field vector and the line of sight. For a
fixed frequency $f$, the relation implies $F(f)\propto
B^{0.9\delta-0.22}$. Hence, for any plausible value of $\delta$, the
optically thin emission increases with increasing magnetic field
strength. In contrast, for optically thick emission at a fixed
frequency, $F(f) \sim \eta_f/\kappa_f \propto T_{eff} \propto
s^{0.5+0.085\delta} (\sin\theta)^{-0.36-0.06\delta}$. Again, for
fixed conditions, $F(f)\propto B^{-0.5-0.085\delta}$ and the
optically thick emission decreases with increasing magnetic field.

\begin{figure}
\epsscale{0.85} \plotone{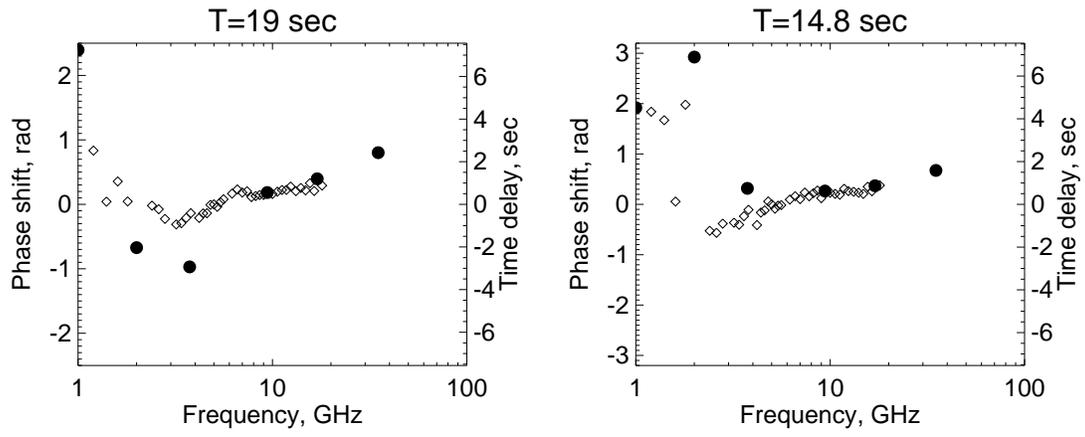}
   \figcaption{\small  Phase
differences  and corresponding time delays  for two main Fourier
peaks. The mean value of the frequency-dependent period (see
Figure~\ref{Fourier_radio}) indicated on top of each panel is taken
to calculate the time delay. \label{phase_delay}}
\end{figure}

\begin{figure}
\epsscale{1} \plotone{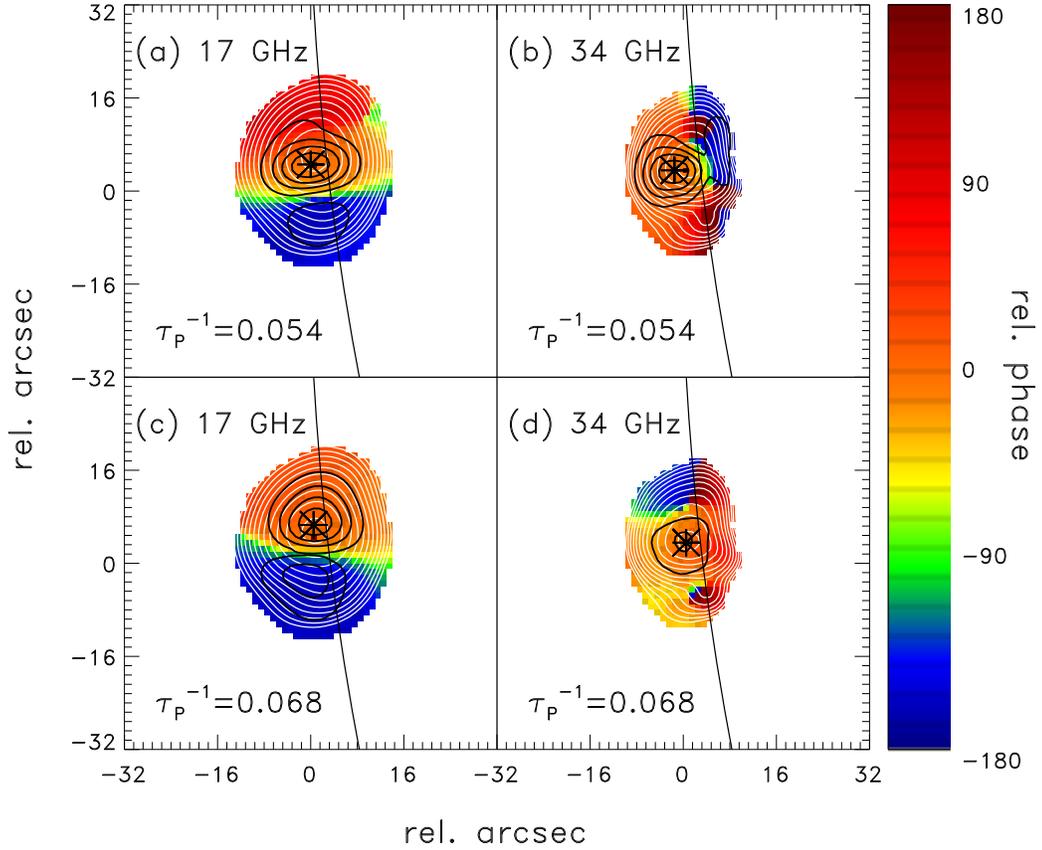} \figcaption{Each panel shows the
Fourier phase (color), amplitude (black contours), and flux density
at the maximum of the flare (white contours) at the radio frequency
and Fourier frequency indicated. The phase in each case is referred
to that at the location of the amplitude maximum, and ranges between
$\pm 180^\circ$ (note that the dark blue and the dark red regions,
showing the phase difference about 360$^\circ$, correspond to
physically the same phase). The amplitude contours are shown at 0.2,
0.4, 0.6, and 0.8 times the maximum in each panel. The white
contours are the same as those shown in Figure~\ref{summary}c (for
panels a and c) and the same as those shown in Figure~\ref{summary}d
(for panels b and d).\label{norh_phases}}
\end{figure}

\begin{figure}
\includegraphics[angle=0,scale=0.85]{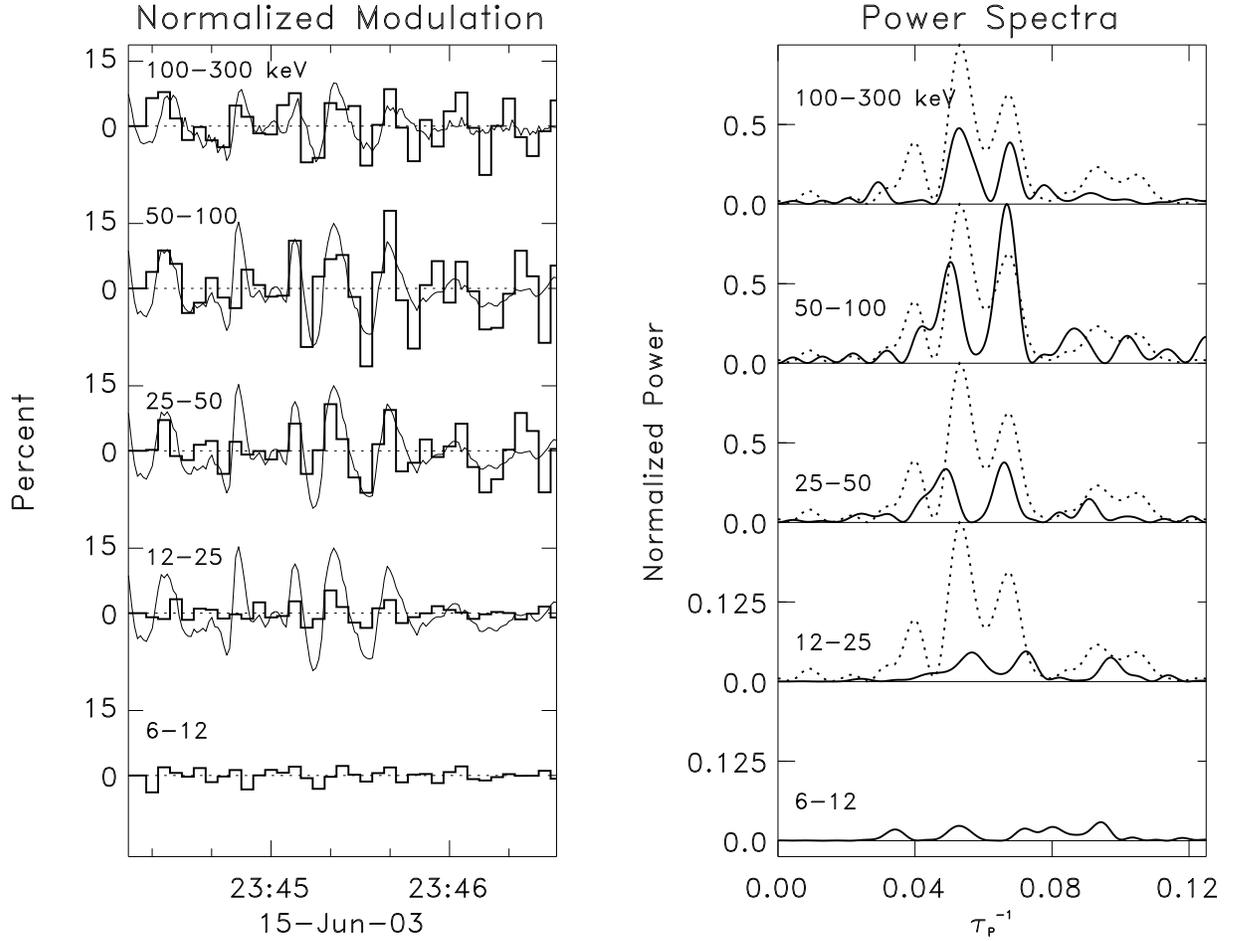}
\figcaption{\small  The normalized modulations of the RHESSI data
(left) and the corresponding power spectra (right). The normalized
modulation of the NoRP 17 GHz emission is  plotted as a thin solid
line over the 12-25, 25-50, 50-100, and 100-300 keV normalized
modulations in the left panel. The 17 GHz power spectrum is  plotted
as a dashed line on the corresponding HXR power spectra.
\label{Fig4}}
\end{figure}

\begin{figure}
\epsscale{0.85} \plotone{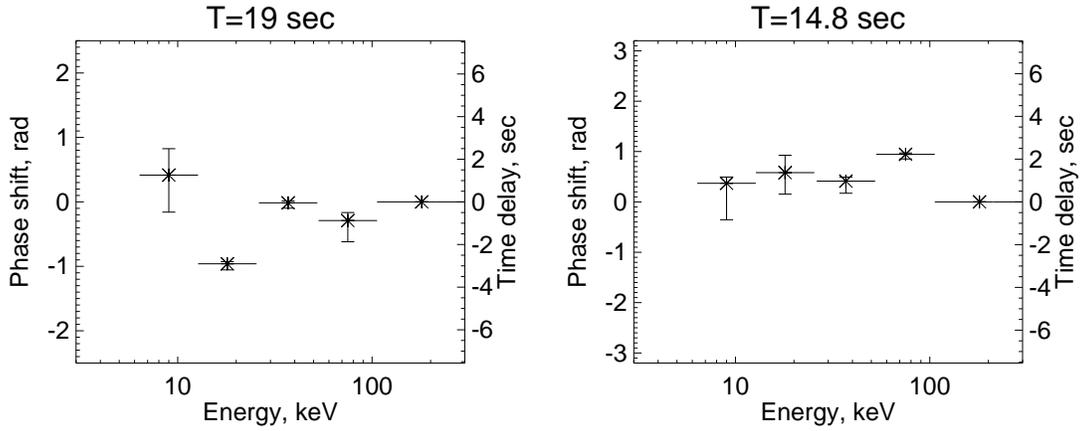}
 \figcaption{\small  Phase differences  and corresponding time delays  for two
main Fourier peaks available in Figure~\ref{Fig4}.  \label{X_fft}}
\end{figure}

\begin{figure}
\epsscale{0.85} \plotone{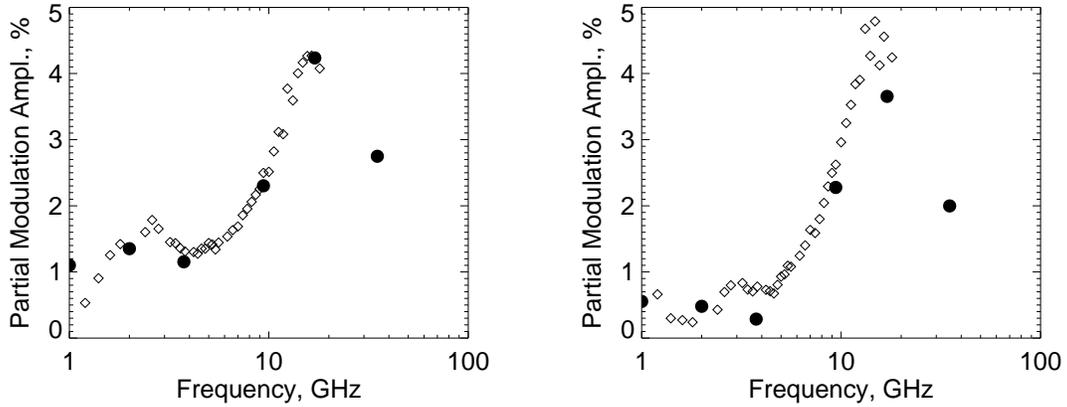}
 \figcaption{\small  Partial
modulation amplitudes related to limited range of Fourier harmonics
around two main Fourier peaks ($\tau_P\approx$19~s, left, and
$\tau_P\approx$15~s, right). \label{periods_radio}}
\end{figure}

\begin{figure}
\epsscale{0.5} \plotone{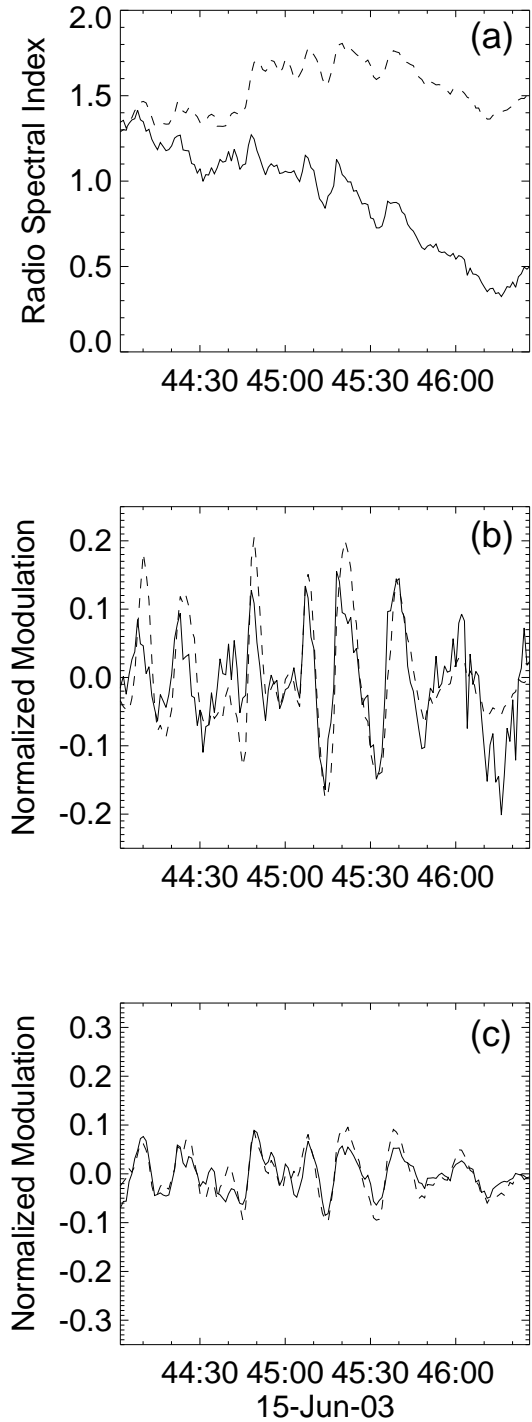}
 \figcaption{\small a) The solid line
shows the spectral index of optically thin emission $\beta_{\rm
thin}$, calculated between 17 and 35 GHz. The dashed line shows the
spectral index of optically thick emission $\beta_{\rm thick}$,
calculated between 3.75 and 9.4 GHz. b) The normalized modulation of
the optically thin spectral index $\beta_{\rm thin}$ compared with
that of the 17 GHz emission. c) The normalized modulation of the
optically thick spectral index $\beta_{\rm thick}$ compared with
that of the 9.4 GHz emission. \label{NoRP_curves}}
\end{figure}

\begin{figure}
\epsscale{0.5} \plotone{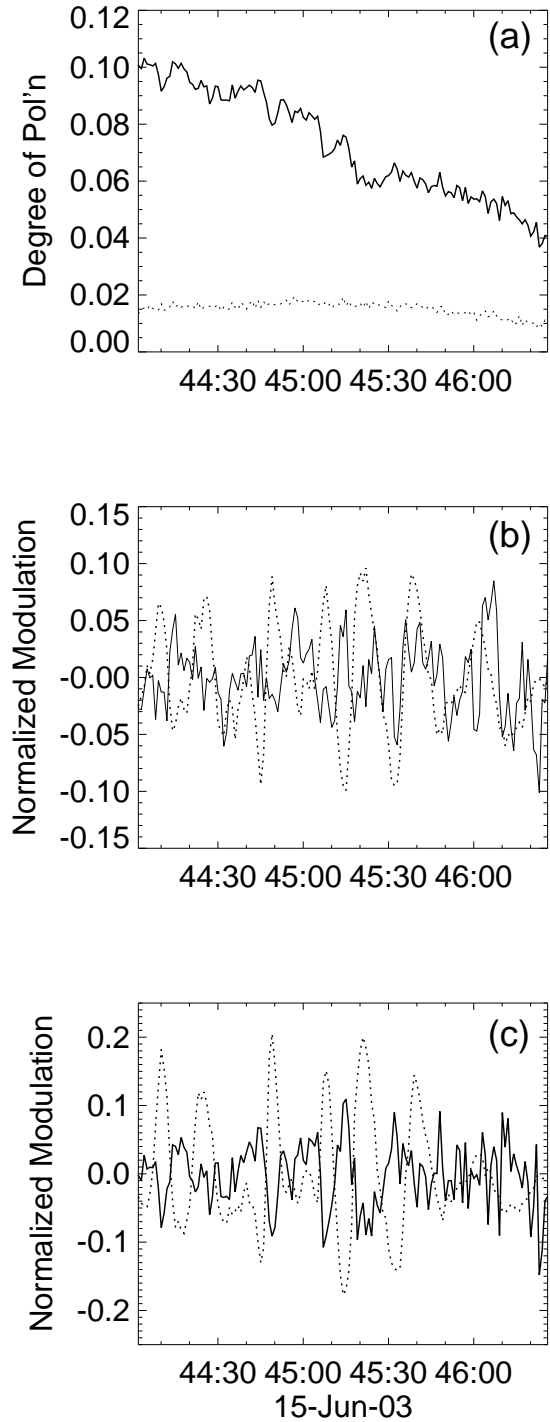}
 \figcaption{\small
a) Degree of polarization of the 17 GHz (solid) and 9.4 GHz (dotted)
emission; b) The normalized modulation of the 9.4 GHz degree of
polarization (solid) compared with that of the 9.4 GHz total flux
(dotted); c) The normalized modulation of the 17 GHz degree of
polarization (solid) compared with that of the 17 GHz total flux
(dotted). \label{Pol_osc}}
\end{figure}

For a global sausage mode oscillation, therefore, variations in the
optically thick emission are expected to be anticorrelated with
those seen in the optically thin emission. The same conclusion can
be drawn for propagating waves in the magnetic loop. However, as
shown in \S2.2.1 and Figure~\ref{phase_delay}, this is not the case.
Instead of a large, 180-degree sharp phase shift from optically
thick to optically thin frequencies, only a slow gradual variation
of oscillation phase is seen. One can easily check that, similar to
the global sausage mode, the low spatial harmonics of the sausage
mode loop oscillations fail to provide a reasonable fit to the phase
variations presented in Figure~\ref{phase_delay} even in the case of
nonuniform magnetic loop.

Consider instead a torsional oscillation which results in a
variation of the magnetic field vector to the line of sight. For
optically thin emission, the dependence of the emissivity on the
angle $\theta$ between the line of sight and the magnetic field
vector is $\eta_f\propto (\sin\theta)^{-0.43+0.65\delta}$ while for
optically thick emission we have $T_{eff}\propto
(\sin\theta)^{-0.36-0.06\delta}$. Again, we expect the optically
thick and optically thin emission to be anti-correlated, which is
not observed.

The polarization data can in principle constrain the underlying
physical model \citep[e.g.,][]{Altyntsev_etal_2008}. However,  the
radio source is located close to the limb so that the polarization
can be strongly affected by propagation effects and, moreover, the
source is not well spatially resolved, see Figure~\ref{summary}.
Therefore, we only rely here on the relative variations and trends
of the degree of polarization rather than on its absolute value. The
degree of circular polarization from optically thin emission is
$\rho_c \propto s^{-0.78+0.55\cos\theta}\propto
B^{0.78-0.55\cos\theta}$ (Dulk 1985) so it is expected to increase
with magnetic field and hence, with the emissivity. Therefore,
variations in $\rho_c$ \textit{should be correlated} with variations
in the flux density in optically thin frequencies if MHD
oscillations are relevant. The observations show, however, that
$\rho_c$ is instead \textit{anticorrelated} with flux variation at
17 GHz (Figure~\ref{Pol_osc}).

Understanding the time variation of the spectral index and its
correlation with radio emission (Figure~\ref{NoRP_curves}) is  also
problematic in the context of MHD oscillations.  While variations in
the magnetic field due to oscillations yield variations in the
(perpendicular component) of the electron energy due to betatron
acceleration, no change is expected in the electron spectral index
$\delta$ of a power-law distribution of electrons \citep[see,
e.g.,][]{Bogachev_Somov_2007} and therefore, no change in the radio
spectral index at optically thin frequencies. However, oscillations
of the magnetic field can give rise to radio spectral index
variations at fixed frequencies because for stronger magnetic field
the spectral peak moves towards higher frequencies, making the
spectrum at fixed frequencies above the peak slightly flatter. In
our event, however, the radio spectrum is steeper when the emission
is stronger.

Finally, consider whether MHD oscillations can provide a reasonable
quantitative fit to the spectral behavior of the modulation
amplitude. To achieve this goal we start from a uniform source
model, whose radio emission is described by simplified Dulk and
Marsh formulae \citep{Dulk_1985}. Although those formulae have a
somewhat limited   range of applicability, they nevertheless provide
useful insights into the parametric dependences and corresponding
trends. Note that the use of the Dulk and Marsh approximation means
that we did not consider the Razin-effect, but rather assume that
the spectral peak is determined by the effect of optical thickness,
which is directly confirmed by the high brightness temperature
determined from imaging radio measurements, (see caption to
Figure~\ref{summary}).

To model an MHD oscillation, we assume sinusoidal oscillations of
the magnetic field at the source
\begin{equation}
\label{B_QPP}
 B= B_{0} + \Delta B \sin\omega t.
\end{equation}
Note, that depending on the specific mode of the magnetic
oscillations, some other parameters of the source (volume, density
etc.) may or may not oscillate. Although we examined a few
particular cases, we present here a model with a simple loop
oscillation (i.e., global sausage mode) taking into account the
conservation of the magnetic flux through the source, which implies:
$S \propto B_0/B$, where $S$ is the source area, and $N_e \propto
B/B_0$. Other models that we have considered yield similar
conclusions.

Specifically, dashed curves in Figure~\ref{model} are calculated for
the uniform model, which are in obvious disagreement with the data.
Although we are able to match a considerable part of the modulation
amplitude curve, the model and observations clearly diverge at low
and high frequencies. A possible reason for this is oversimplified
model relying on an uniform radio source. Indeed, we have clear
observational evidence of the source inhomogeneity: the observed
optically thick spectrum is much flatter ($\sim f^{1.5}$) than the
theoretical one for the uniform source ($\sim f^{2.5-3}$).

Accordingly, to improve the spectral fit to the data, we introduce a
simplified non-uniform model, which takes into account the source
inhomogeneity assuming slower (than for the uniform case) increase
of the gyrosynchrotron optical depth with frequency decrease to
match the observed low-frequency slope of the spectrum.
Specifically, we adopted $\tau=\tau_0 (f/9~\hbox{GHz})^{1.1}$, where
$\tau_0$ is the optical depth for the uniform source
\citep{Dulk_Marsh_1982}. To justify this simplified non-uniform
model we note that in the optically thick case the emission comes
from a thin layer having the optical depth of about 1 in both
uniform and inhomogeneous cases. In the inhomogeneous case, however,
the physical size of this layer can decrease more slowly than in the
uniform case, giving rise to a flatter low-frequency spectrum, which
is taken into account within our simplified model.

The corresponding model (dash-dotted curves) gives a reasonable
spectral fit, but does not at all improve the fit of the modulation
amplitude vs frequency. Adding some portion of unrelated noise to
the model improves the fit slightly. The model providing the best
fit (solid lines) to the observed modulation amplitude yields
$B_{0}\approx540$ G and $\Delta B/B_{0} \approx 0.06$ and a low
level of the unrelated noise 1.7$\%$. However, if we tweak the model
for the best fit to the modulation amplitude data (solid line), the
model gives a poor fit to the power spectrum. In particular, the
model spectral peak frequency is significantly lower than the
observed one. In contrast, if we match the observed spectrum
(dash-dotted line), we cannot get a reasonable fit to the modulation
amplitude. Overall, irrespectively to the details adopted, the model
of the MHD oscillation strongly overestimates the modulation
amplitude at low and high frequencies.

We conclude that while the periods of the QPPs at HXR and radio
wavelengths are consistent with MHD modes that could be supported by
a coronal magnetic loop, a more detailed analysis of the
observations shows that MHD oscillations fail to account for most of
the observed properties of the QPPs.   We conclude that  MHD loop
oscillations are not relevant to the flare studied here.

\subsection{QPPs as a Result of Electron Acceleration and Injection}

We now consider whether quasi-periodic electron acceleration and
injection can satisfactorily account for the observations. In
particular, we suggest that each peak in the radio and HXR emission
corresponds to a discrete injection of fresh 
electrons into the magnetic loop. The modulation amplitude of the
radio emission, which is comparable with that of HXR emission, as
well as the short decay time of the radio pulses, suggest that the
newly injected fast electrons have a rather short life time, a few
seconds. It is unlikely that the short electron life time is due to
Coulomb losses in the loop, since this would require a rather high
plasma density $n>3\times 10^{11}$~cm$^{-3}$ that would, in turn,
result in strong Razin suppression of the radio emission at low
frequencies and also in a spectral peak well above 10 GHz, which are
not observed. Therefore, the most probable reason of the rapid loss
of the energetic electrons is emptying of the loss-cone by
precipitation. This idea, corresponding to the DPTPP model, has
several consequences.

First, the correlation between the radio and HXR emission is easily
explained. With each new injection of 
electrons into the trap, the optically thin emission increases as
$F(f)\propto N$ and the electrons filling the loss cone directly
precipitate from the loop, producing enhanced HXR emission. Those
electrons with large enough pitch angles accumulate in the trap,
where their energy distribution is expected to collisionally harden.

In contrast to the case of MHD oscillations, the phase of the
Fourier components is not expected to change between optically thick
and thin emission. Moreover, the amplitude of the oscillation is
expected to greatly diminish as emission transitions from optically
thin to optically thick radio emission. The observations are
qualitatively consistent with these expectations.  This alone is
sufficient to argue strongly for quasi-periodic injection of
particles as the cause of the oscillations.  However, the
observations are sufficiently complete that we can investigate the
oscillations in greater detail. Specifically, the phase of the
dominant Fourier peak varies systematically from high to low
frequencies, corresponding to a relative timing delay between
frequencies. The sense of the delay is such that high
frequencies/energies lag low frequencies/energies. Since the radio
frequencies/X-ray energies are, broadly speaking, proportional to
electron energy, these observations imply that the acceleration and
injection of higher energy electrons in each oscillation lag the
lower energy electrons. It automatically follows that the spectral
index of the injected electrons must be modulated.

These ideas can be captured in a simplified phenomenological model
of the periodic injection of energetic electrons into the magnetic
loop.  The number density of nonthermal electrons $N_e$ is taken to
be a superposition of periodically injected electrons $\Delta N_e$
and those electrons that have accumulated in the magnetic trap
$N_{e0}$:
\begin{equation}
\label{Ne_QPP}
 N_e= N_{e0} + \Delta N_e |\sin\omega t |.
\end{equation}

In addition, we assume that the electron spectral index at the
source varies almost synchronously with the number density, although
with a small phase shift, which agrees with that in
Figure~\ref{NoRP_curves}:

\begin{equation}
\label{del_QPP}
 \delta= \delta_{0} + \Delta \delta |\sin(\omega t+\varphi_{\delta})|.
\end{equation}

The relative variation of the spectral index, $\Delta
\delta/\delta=0.08$, and the phase $\varphi_{\delta}\approx\pi/16$
are taken to provide a change in radio spectral index $\Delta
\beta/\beta \approx 0.08$ and the corresponding time shift that
agrees with Figure~\ref{NoRP_curves}b, while the value $\Delta
N_e(>100~keV)/N_e(>100~keV)\approx0.6$ is selected to match the
modulation amplitude in the range of the radio spectral peak. The
magnetic field at the source is taken to be $B= 650$ G, which
provides the correct radio flux density and spectral peak frequency
$f_{\rm peak}$   near $\sim\!10$~GHz.

\begin{figure}
\epsscale{0.85} \plotone{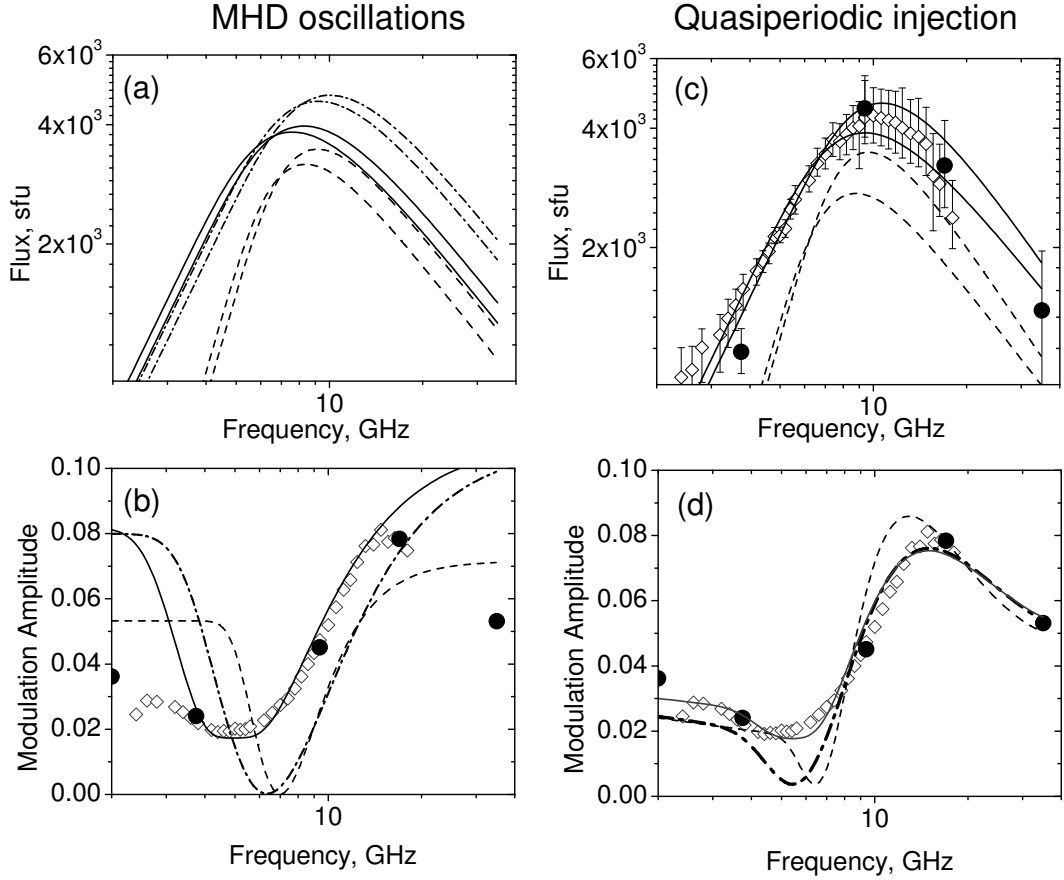}
 \figcaption{\small  Model gyrosynchrotron spectra
(the pairs show two extremes as the spectrum oscillates for each
case) and modulation amplitudes for the cases of MHD global sausage
mode oscillations (left) and quasiperiodic injection (right). Dashed
curves are calculated for a uniform model, while the dash-dotted and
solid curves for simplified inhomogeneous models. In the left column
the non-uniform spectra differ from each other: dash-dotted curves
provide good spectrum fit (a), but fail to account the modulation
amplitude (b), while solid curves suggest the best fit to the
modulation amplitude (b), but fail to provide a reasonable spectrum
fit (a). The solid curve in the  panel (d), providing an excellent
fit to the data, is calculated for the inhomogeneous model with an
unrelated noise contribution added with the frequency-independent
rms value of 1.7$\%$. The corresponding solid and dash-dotted
spectra are undistinguishable in the panel (c). The diamonds show
the OVSA data, while the filled circles show the NoRP data. A
composite observed spectrum (diamonds and circles) displayed in
panel (c) shows the mean values of the radio flux as observed for
the first 80 s of the analyzed data fragments, while the
corresponding error bars display the whole range of the flux
variation during these 80 s. We do not show this spectrum in panel
(a) to keep it clearly readable. \label{model}}
\end{figure}

Figure~\ref{model}, right column, displays the spectra and the
modulation amplitude for the quasi-periodic injection model along
with the respective measurements. As in the previous section, dashed
curves in Figure~\ref{model} are calculated for a uniform model, and
are in qualitative agreement with the observed modulation amplitude.
However, the fit to the observed spectrum can be improved using a
non-uniform model, as has been explained in the previous section.

The simplified non-uniform model simultaneously gives a reasonable
spectral fit (dash-dotted curves) and a better approximation to the
modulation amplitude vs. frequency. The remaining discrepancy near
the minimum of the curve is surprisingly low.  It can be adjusted by
adding 1.7$\%$ unrelated noise as was also necessary for the MHD
model in  Figure~\ref{model}b.  This level is only about 20-30 sfu
at the frequencies near the minimum. Such a small residual
oscillation does not affect the spectrum itself
(Figure~\ref{model}c). The corresponding (solid) curve provides an
excellent fit to the modulation amplitude data (Figure~\ref{model}d)
showing that the simplified model can account for the observations
assuming only quasi-periodic injection of the fast electrons into
the radio source. Although we fit here the total modulation
amplitude of the radio emission, we note that all partial modulation
amplitudes (including those given in Figure~\ref{periods_radio})
behave qualitatively similarly, therefore, all oscillations found in
the radio data should be ascribed to the same process.

To account for the observed anti-correlation between flux density
variations and polarization (Figure~\ref{Pol_osc}a), we find a
natural explanation in a systematic variation in pitch-angle
anisotropy of the electrons. \cite{FlMel_2003b} have shown that the
degree of polarization of optically thin gyrosynchrotron emission at
a fixed frequency increases as the degree of perpendicular
anisotropy in the electron distribution function increases. Assuming
for simplicity that the electrons are injected with an isotropic
distribution, after some time the loss cone will empty due to escape
of electrons from the trap, and the pitch angle distribution of the
remaining, trapped electrons becomes anisotropic, leading to an
increase in polarization.

To summarize, we find that the observed properties of the QPPs in
the 15 June 2003 flare are consistent with the quasi-periodic
injection of energetic electrons into a coronal magnetic loop. The
injected electrons show an evolution wherein more energetic
electrons are delayed relative to less energetic electrons. This
manifests itself in a quasi-periodic modulation of the index
$\delta$. With each fresh injection of electrons, those with small
pitch angles directly precipitate from the trap, producing HXR
emission. The distribution of those electrons that remain trapped
collisionally hardens \citep{Melnikov_1994}, yielding systematic
trends in the radio spectral index $\beta_{\rm thin}$ and $\rho_c$,
both of which decrease in time for optically thin emission. All of
the essential features of the radio and HXR QPPs are consistent with
the DPTPP picture.

\section{Discussion}

In this paper we present a detailed quantitative analysis of the
radio and HXR QPPs in a solar flare and   fit the data assuming two
competing models --   one model involving MHD loop oscillations and
the  other model involving quasiperiodic injection/acceleration of
nonthermal electrons. We find, using the complete spectral
information available for this event, that it is not possible to get
a consistent fit to the data within the magnetic oscillation model.
By comparison, the model with quasiperiodic particle injection
offers an excellent fit to the variations in radio flux density,
power-law index, and polarization, as well as their correlations
with each other and with hard X-ray QPPs.

Although in principle one might expect a loop with quasi-periodic
injection to respond with MHD oscillations of some type, our
analysis of partial modulation amplitude suggests that all of the
significant periods (peaks of the Fourier spectrum) are due to
quasi-periodic injection of the fast electrons into the source. This
fact implies that the magnetic loop comprising the radio source
either is a rather bad resonator or has no eigen-mode in the
considered range of the oscillation periods (in essence, above 0.2
s). Indeed, if the loop could support the corresponding
oscillations, it would respond on the quasi-periodic particle
injection by displaying an appropriate eigen-mode of the loop
oscillation, which is not observed.

The  modulation amplitude of the gyrosynchrotron (GS) emission,
which is comparable with that of HXR emission, as well as the short
decay time of the radio pulses, suggest that the newly injected fast
electrons have a rather small life time ($\lsim$  few s). That short
life time can hardly be provided by the Coulomb losses in the loop,
since this would require a high plasma density $n>3\times
10^{11}$cm$^{-3}$ resulting in strong Razin suppression, which we
have argued is inconsistent with the observations. Therefore, the
most probable reason for the rapid loss of the energetic electrons
is emptying of the loss-cone by precipitation, which also accounts
for  the excellent degree of correlation between the radio and HXR
QPPs and the polarization properties of the event.

Finally, we note that this event is similar in many respects to the
celebrated flare of 7 June 1980, discussed in great detail by Kane
et al. (1983). In particular, Kane et al. also reported an
anti-correlation between the degree of polarization and the flux
density at 17 GHz. Because the spectral peak of the emission in that
event was $\gtrsim17$ GHz, so that optical depth effects may be
important, we choose 9.4 GHz in our event for a direct comparison.
The 9.4 GHz emission for the 15 Jun 2003 event, like the 17 GHz
emission for the 7 Jun 1980 flare, is marginally optically thick.
The normalized modulations of the intensity and polarization at 9.4
GHz in our event are indeed anti-correlated (see
Figure~\ref{Pol_osc}), and similar in value to those at 17 GHz in
the  1983 event.  Although Kane et al. do not suggest an
interpretation of the polarization oscillations, they conclude that
quasiperiodic electron acceleration and injection are responsible
for the 1983 event. We find that the variation of the angular
particle distribution during the quasiperiodic electron injection
results naturally in the oscillations of the degree of polarization,
as observed in both Kane's and our events, which eventually confirms
the Kane et al. choice in favor of the quasiperiodic
injection/acceleration in that event.

\section{Conclusions}

We described an oscillating event observed with high spectral
resolution in the microwave range and analyzed it by applying and
developing further the Fourier method presented earlier by
\cite{PRL,Fl_etal_2002,Melnikov_etal_2005}. We developed the idea of
partial modulation amplitudes, expressed as the modulation amplitude
in each particular Fourier peak of finite bandwidth, using the
Parseval identity. In the general case the method is capable of
distinguishing different simultaneous contributions to the overall
modulation of the radio emission, which might be especially powerful
when both MHD oscillations and quasiperiodic electron injection
contribute to QPPs of the radio emission. For our event this method
showed that all of the Fourier harmonics have the same
characteristics, i.e. are all due to quasiperiodic injection of the
electrons in the radio source. Thus, we conclude that for this event
MHD oscillations play no role. Possibly, this result calls into
question some previous studies, which concluded that MHD
oscillations play a key role in the radio and HXR QPPs, but which
did not have the complete spectral information needed to apply our
method of analysis.

Specifically, QPPs in our event had the following characteristics:

$\bullet$ The Fourier spectrum at each observing frequency is composed of a few
significant peaks, corresponding to a few oscillation quasi-periods.

$\bullet$ None of the main radio Fourier peaks could be interpreted
as the result of MHD loop oscillations.

$\bullet$ Several measures, e.g., modulation amplitudes of the flux
density as a function of radio frequency or photon energy, the
relative phases of the oscillations as a function of frequency, the
spectral indices of the radio emission, and the degree of
polarization  are   inconsistent with MHD loop oscillations, but are
fully consistent with quasi-periodic acceleration and injection of
electrons as the cause of the  observed radio and X-ray QPPs.

$\bullet$ Rather strong QP energy release and particle injection
appears not to have excited MHD oscillations, hence, the
corresponding loop is a rather bad resonator in the range of
oscillation frequencies under discussion.

The spatial resolution of the available data is insufficient to
address the physical cause of the observed quasiperiodic injection
of fast electrons. Several mechanisms have been advanced in the
literature, such as a nonlinear self-organizing regime of the
electron acceleration \citep[e.g.,][]{Aschwanden_1987} or bursty
reconnection  during a plasmoid ejection \citep{kliem_etal_2000} or
during the interaction of current carrying loops
\citep{Sakai_Ohsawa_1987}. In any case, acceleration of fast
electrons in the form of distinct, quasi-periodic injections seems
to be rather common in solar flares, see, e.g.,
\citep{Aschwanden_etal_1998, Benz_etal_2005, Altyntsev_etal_2008}.
In many events, the accumulation of the electrons in the magnetic
loop due to the trapping effect results in smooth light curves,
making it difficult to distinguish the contributions from these
distinct injections. In some favorable conditions, however, when
these injections are comparable in strength, more or less
equidistant in time, and the fraction of directly precipitating
electrons is relatively large, such repetitive electron injections
manifest themselves as pronounced broadband pulsations of the radio
emission, as in the instance considered in this paper.

We conclude that radio and X-ray spectral data, when available,
should be used along with imaging observations to firmly determine
whether QPPs are produced by quasiperiodic injection of fast
electrons into a magnetic loop or some MHD oscillating mode of the
loop. The method of partial modulation amplitudes should even be
able to distinguish cases when both processes are operating
simultaneously.

\acknowledgments We thank Gordon Hurford for his careful calibration
of the RHESSI data used in this work. We appreciate valuable
comments to the paper draft made by Victor Grechnev. The National
Radio Astronomy Observatory is a facility of the National Science
Foundation operated under cooperative agreement by Associated
Universities, Inc. This work was supported in part by NSF grants
AST-0607544 and ATM-0707319 and NASA grant NNG06GJ40G to New Jersey
Institute of Technology, and by the Russian Foundation for Basic
Research, grants No. 06-02-16295, 06-02-16859, 06-02-39029. We have
made use of NASA's Astrophysics Data System Abstract Service.

\clearpage

\end{document}